\documentclass[twocolumn,amsmath,amssymb]{revtex4}
\usepackage{graphicx}
\usepackage{dcolumn}
\usepackage{bm}
\begin{document}

\title{Controllable, driven phase transitions in the Fractional quantum Hall 
states in bilayer graphene}
\author{Vadim M. Apalkov}
\affiliation{Department of Physics and Astronomy, Georgia State University,
Atlanta, Georgia 30303, USA}
\author{Tapash Chakraborty$^\ddag$}
\affiliation{Department of Physics and Astronomy,
University of Manitoba, Winnipeg, Canada R3T 2N2}

\date{\today}
\begin{abstract}
Here we report from our theoretical studies that in biased bilayer graphene,
one can induce phase transitions from an incompressible fractional quantum 
Hall state to a compressible state by tuning the bandgap at a given electron 
density. The nature of such phase transitions is different for weak and 
strong inter-layer coupling. Although for strong coupling more levels 
interact there are lesser number of transitions than for the weak coupling 
case. The intriguing scenario of tunable phase transitions in the fractional 
quantum Hall states is unique to bilayer graphene and never before existed in 
conventional semiconductor systems.
\end{abstract}
\maketitle

The unconventional quantum Hall effect in monolayer graphene, whose experimental
observation \cite{novoselov_kim} unleashed quite unprecedented interest in this
system \cite{review}, reflects the unique behavior of massless Dirac fermions in
a magnetic field \cite{wallace,mcclure}. In bilayer graphene this effect confirms
the presence of massive chiral quasiparticles \cite{novoselov_bi}. An important
characteristic of bilayer graphene is that it is a semiconductor with a tunable
bandgap between the valence and conduction bands \cite{pereira}. This modifies 
the Landau level spectrum and influences the role of long-range Coulomb 
interactions \cite{abergel}. Here we report that the fractional quantum Hall 
effect (FQHE), a distinct signature of interacting electrons in the system 
\cite{fqhe_book,stormer} is very sensitive to the interlayer coupling strength 
and the bias voltage. We propose that by tuning the bias voltage one can induce 
phase transitions from an incompressible state to a compressible state at a 
given gate voltage. The bilayer graphene system shows quite different properties 
for weak and strong inter-layer coupling. For a weak coupling the energy spectrum 
as a function of bias voltage shows a set of anti-crossings, resulting in 
transitions from the FQHE state to a compressible state. At strong coupling 
there is a strong interaction between many energy levels, which finally results 
in only a few phase transitions. This interesting scenario of tunable phase 
transitions in the FQH states is unique to bilayer graphene. In conventional 
semiconductor systems the type of phase transitions discussed below was
never reported. The FQHE in monolayer graphene was in fact, studied theoretically 
by us \cite{vadim_fqhe} and subsequent experiments confirmed the existence 
of that effect in suspended monolayer graphene samples \cite{fqhe_kim}. 
No such studies have been reported on bilayer graphene.

We assume that the bilayer graphene consists of two coupled graphene layers with
the Bernal stacking arrangement. Our main concern then is the coupling between 
atoms of sublattice A of the lower layer and atoms of sublattice B$^{\prime}$ of 
the upper layer. The single-particle levels have two-fold spin degeneracy and 
two-fold valley degeneracy, which can be lifted in the many-particle systems at 
relatively large magnetic fields \cite{nakamura}. The valley degeneracy is also 
lifted under an applied bias voltage \cite{novoselov_bi}. Considering only one spin
direction, we describe the state of the system in terms of the four-component 
spinor $(\psi^{}_A, \psi^{}_B, \psi_{B^{\prime}}, \psi_{A^{\prime}})^T$ for valley $K$ 
and $(\psi_{B^{\prime}}, \psi_{A^{\prime}}, \psi^{}_A, \psi_B)^T$ for valley $K^{\prime}$. 
Here subindices A, B and A$^{\prime}$, B$^{\prime}$ correspond to lower and upper 
layers respectively. The strength of inter-layer coupling is described in terms 
of the inter-layer hopping integral, $t$. In a biased bilayer graphene the bias 
potential is introduced as the potential difference, $\Delta U$, between the upper 
and lower layers. The Hamiltonian of the biased bilayer system in a perpendicular 
magnetic field then takes the form \cite{novoselov_bi}
\begin{equation}
{\cal H} =  \xi \left(
\begin{array}{cccc}
    \Delta U/2 & v^{}_F \pi^{}_+ & \xi t & 0   \\
    v^{}_F \pi^{}_- & \Delta U/2 & 0 & 0 \\
    \xi t & 0& -\Delta U/2 & v^{}_F \pi^{}_-   \\
    0 & 0&  v^{}_F \pi^{}_+ & -\Delta U/2
\end{array}
\right),
\label{H1}
\end{equation}
where $\pi^{}_{\pm}=\pi^{}_x\pm i\pi^{}_y$, $\vec{\pi}=\vec{p}+e\vec{A}/c$,
$\vec{p}$ is the two-dimensional electron momentum, $\vec{A}$ is the vector 
potential, $v^{}_F \approx 10^6$ m/s is the fermi velocity, and $\xi =+$ ($K$ 
valley) or $-$ ($K^{\prime}$ valley).

In a perpendicular magnetic field the Hamiltonian (\ref{H1}) generates a discrete
Landau level energy spectrum. The corresponding eigenfunctions can be expressed in
terms of the conventional nonrelativistic Landau functions. The electron states
in sublattices A and A$^{\prime}$ are written in terms of the $n$-th Landau functions, 
while the electron states in sublattices B and B$^{\prime}$ are described by the 
$|n-1|$ and $n+1$ Landau functions, respectively. Therefore the Landau states in 
bilayer graphene can be described as a mixture of $n$, $n+1$, and $n-1$ 
nonrelativistic Landau functions belonging to different sublattices \cite{pereira}. 
This mixture, for a given value of $n$, results in four different Landau levels. 
The Landau level energies, $\varepsilon$ corresponding to the 
index $n$ can be found from the following equation \cite{pereira}
\begin{equation}
\left[\left(\varepsilon + \xi \delta\right)^2 - 2(n+1)\right]
\!\!\! \left[(\varepsilon - \xi \delta )^2 - 2n \right] = (\varepsilon ^2 - \delta ^2 )t^2 ,
\label{level1}
\end{equation}
where $\delta = \Delta U/2$ and all energies are expressed in units of $\hbar v^{}_F/\ell^{}_0$.
Here $\ell^{}_0 = (\hbar/eB)^{\frac12}$ is the magnetic length.

We now introduce a labeling scheme for Landau levels in bilayer graphene. From 
Eq.~(\ref{level1}), we see that for each value of $n$ ($=0,1,2,\ldots $) and 
in each valley there are four solutions, i.e., four Landau levels. Usually, the two 
lower Landau levels have negative energies and the two upper Landau levels have 
positive energies. Then each of the four Landau levels can be labeled as $n^{(\xi)}_i$, 
where $i=-2,-1,1,2$ is the label of the Landau level corresponding to the solution 
of Eq.~(\ref{level1}) for a given value of $n$ in the ascending order.

For a partially occupied Landau level the properties of the system, e.g., the
ground state and excitations, are completely determined by the inter-electron
interactions, which can be expressed by Haldane's pseudopotentials, $V_m$,
\cite{haldane} (energies of two electrons with relative angular momentum $m$). 
In a graphene bilayer the Haldane pseudopotentials in a Landau
level with index $n$ and the energy $\varepsilon$ have the form
\begin{equation}
V_m^{(n)} = \int _0^{\infty } \frac{dq}{2\pi} q V(q)
\left[F_{n, \varepsilon } (q) \right]^2 L_m (q^2) e^{-q^2},
\label{Vm}
\end{equation}
where $L_m(x)$ are the Laguerre polynomials, $V(q) = 2\pi e^2/(\kappa \ell^{}_0 q)$
is the Coulomb interaction in the momentum space, $\kappa$ is the
dielectric constant, and $F_{n,\varepsilon} (q)$ are the corresponding form
factors
\begin{eqnarray}
F_{n,\varepsilon}(q) = &  &
\frac{1}{d^{}_n} \left[
\left(1+ f_n^2  \right) L_n\left(\tfrac{q^2}2\right) +
\frac{2n}{(\varepsilon - \xi \delta )^2} L_{n-1}\left(\tfrac{q^2}2\right)\right.
\nonumber \\
 & & +\left. \frac{2(n+1)}{(\varepsilon + \xi \delta)^2} f_n^2 L_{n+1}\left(
\tfrac{q^2}2\right)\right],
\label{fn}
\end{eqnarray}
where $f^{}_n = \frac{(\varepsilon -\xi \delta)^2 -2n}{t(\varepsilon - \xi \delta)}$ and
$d^{}_n =1+ f_n^2 + \frac{2n}{(\varepsilon - \xi \delta)^2} +
\frac{2(n+1)}{(\varepsilon + \xi \delta)^2} f_n^2.$

The form factors of bilayer graphene [Eq.~(\ref{fn})] are clearly different
from the corresponding ones for a monolayer graphene. In the latter case,
the form factor of the $n=0$ Landau level is the same as that of conventional 
non-relativistic electrons \cite{vadim_fqhe,goerbig}, $F_{0}(q)=L_0 $.
The form factors of higher Landau levels are determined by the mixture of
$L_n$ and $L_{n+1}$ terms. In bilayer graphene the form factors of the
$n=0$ Landau level are mixtures of the $L_0$ and $L_1$ terms and are different 
from that in the non-relativistic case. There is one special Landau level in 
bilayer graphene with index $n=0$, whose properties are completely identical to 
that of the non-relativistic $n=0$ Landau level. It is clear from Eq.~(\ref{level1}) 
that for $n=0$ there is a Landau level with energy $\epsilon = \xi \delta$. This
energy does not depend on the coupling between the layers, 
$t$. The form factor of this Landau level is exactly equal to the form factor of 
a non-relativistic system of the $n=0$ Landau level, $F_{n=0,\epsilon = \xi \delta} 
= L_0$. Therefore, all many-body properties of a bilayer system in the $n=0$, $\epsilon =
\xi \delta $ Landau level are completely identical to those of a non-relativistic
conventional system in the $n=0$ Landau level.

For Landau levels with higher indices, the form factor is a mixture
of three different functions, $L_n$, $L_{n-1}$, and $L_{n+1}$. 
Therefore, in general, the strength of inter-electron interactions 
in bilayer graphene is strongly modified as compared to its value in 
monolayer graphene. To address the effects of these modifications on 
the properties of the many-electron system in bilayer graphene we 
investigate fractional filling factors corresponding to the FQHE 
\cite{fqhe_book}. We treat the many-electron system at various fractional 
filling factors numerically within the spherical geometry \cite{vadim_fqhe,haldane}. 
The radius of our spehere is $R=\sqrt{S}\ell^{}_0$, where $2S$ is the number 
of magnetic fluxes through the sphere in units of the flux quanta. The 
single-electron states are characterized by the angular momentum, $S$, 
and its $z$ component, $S_z$. For the many-electron system the corresponding 
states are classified by the total angular momentum $L$ and its $z$ component, 
while the energy of the state depends only on $L$ \cite{fano}. A given 
fractional filling of the Landau level is determined by a special relation 
between the number of electrons $N$ and the radius of the sphere
$R$. For example, the $\frac13$-FQHE state is realized at $S=(\frac32)(N-1)$,
while the $\frac25$-FQHE state corresponds to the relation $S=(\frac54)N-2$.
With the Haldane pseudopotentials [Eq.~(\ref{Vm})] we determine the interaction
Hamiltonian matrix \cite{fano} and then calculate a few lowest eigenvalues
and eigenvectors of this matrix. The FQHE states are obtained when the
ground state of the system is an incompressible liquid, the energy spectrum
of which has a finite many-body gap \cite{fqhe_book,stormer}.

\begin{figure}
\begin{center}\includegraphics[width=7.0cm]{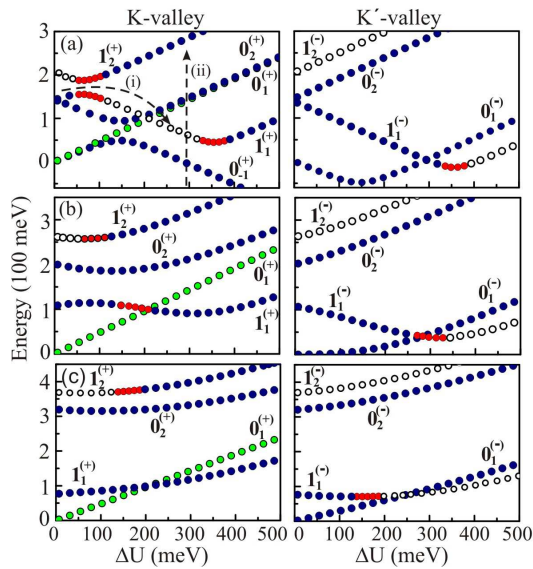}\end{center}
\vspace*{-0.5 cm}
\caption{
A few lowest Landau levels of the conduction band (for two valleys) 
as a function of the bias potential, $\Delta U$, for different values 
of inter-layer coupling: (a) $t=30$ meV (b) $t=150$ meV and (c) $t=300$
meV and a magnetic field of 15 Tesla. The numbers next to the curves 
denote the corresponding Landau levels. The Landau levels where the 
$\frac13$-FQHE can be observed are drawn as blue and green filled dots. 
The green dots correspond to the Landau levels where the FQHE states 
are identical to that of a monolayer of graphene. The red dots represent 
Landau levels with weak $\frac13$-FQHE and the open dots for those where 
the FQHE is absent. In (a), the dashed lines labeled by (i) and (ii)
illustrate two situations: (i) under a constant gate voltage and
variable bias potential; (ii) under a constant bias potential and
variable gate voltage. 
}
\label{figone}
\end{figure}

We begin with the celebrated $\frac13$-FQHE \cite{stormer}, corresponding
to the filling factor $\nu =\frac13$. The behavior of the Landau level 
spectra as a function of the bias voltage and for different values of $t$ 
are displayed in Fig.~\ref{figone} (only the Landau levels with positive 
energies are shown). A similar behavior is valid for other FQHE filling 
factors, e.g., for $\nu=\frac25$ \cite{vadim_unpub}. Figure~\ref{figone} 
clearly illustrates that the FQHE can be observed in all $n=0$ Landau levels 
with the strongest FQHE being in the second $n=0$ Landau levels of both 
valleys, i.e., $0^{(+)}_2$ and $0^{(-)}_2$.

We found an interesting behavior in the $n=1$ Landau levels. There are
four such levels with positive energy; two per each valley.  The FQHE in
these levels shows different properties depending on the strength of 
$t$. For all parameters of the system there is no FQHE in the Landau level 
$1_{2}^{(-)}$. At small values of $t$, $t\lesssim 150$ meV, (Fig.~\ref{figone}(a)) 
the system clearly shows few {\em anti-clossings} accompanied by the 
transitions from the FQHE incompressible state to a state without 
the FQHE. There is one such transition for levels $1_2^{(+)}$ and $1_1^{(-)}$, 
but there are two transitions in level $1_1^{(+)}$, corresponding to two 
anti-crossings in this level. Thus, in level $1_1^{(+)}$ and small 
$\Delta U$, the FQHE is present but disappears at larger values of 
$\Delta U$. It reappears at very large values of $\Delta U (\approx 400$ meV).
With increasing $t$ the two anti-crossings in level $1_1^{(+)}$ merge
(see Fig.~\ref{figone}(b)) and finally disappear (Fig.~\ref{figone}(c)). At 
large values of $t$,  $t > 150$ meV, there are only two anti-crossings 
(Fig.~\ref{figone}(c)) in $1_1^{(-)}$ and $1_2^{(+)}$ Landau levels. At 
such large values of $t$, the anti-crossings cannot be considered as interaction 
between two `crossing' levels, but as a result of strong interaction between 
all (four) levels of the two layers of bilayer graphene. It is important that 
such strong interaction between the levels does not destroy the FQHE, but 
shows well-defined regions with strong FQHE. For weak coupling between graphene
monolayers, i.e. for a small $t$, transitions from the $\frac13$-FQHE state 
to a non-FQHE state can be understood in terms of the anticrossing of
$n=1$ and $n=2$ Landau levels of the monolayers. For monolayers, the FQHE 
can be observed only in the $n=0$ and 1 Landau levels but not for $n=2$ 
\cite{vadim_fqhe}. The levels without the FQHE in Fig.~\ref{figone} (c) 
then correspond to $n=2$ of one of the monolayers. For large $t$, i.e. for 
strong coupling, such a simple description is however inadequate. The 
properties of the $n=1$ levels have important implications for possible 
experimental observations of this unique behavior (Fig.~\ref{figone}(a)):

(i) By applying a gate voltage the electron density can be tuned so that
the first four Landau levels are completely occupied and the next Landau 
level is partially occupied with the FQHE filling factor, for example, 
$\nu =\frac13$. Following Fig.~\ref{figone}(a), this means that the $0_{-1}^{(+)}$, 
$0_{1}^{(+)}$, $1_{1}^{(-)}$, $0_{2}^{(+)}$ Landau levels are fully occupied,
while the $1_{1}^{(+)}$ Landau level has a filling factor $\frac13$. Then, by
varying $\Delta U$ from a small value, e.g., 10 meV, to a larger value, e.g.,
200 meV, one can observe the disappearance of the FQHE (line (i) in 
Fig.~\ref{figone}(a)).

(ii) The bias voltage is kept fixed at a large value, e.g., $\Delta U=300$ 
meV. Then by varying the gate voltage and thus increasing the electron 
density, one can observe the disappearance and reappearance of the 1/3-FQHE 
in higher Landau levels (line (ii) in Fig.~\ref{figone}(a)), when the filling 
factors of the corresponding Landau levels are 1/3.

\begin{figure}
\begin{center}\includegraphics[width=8.4cm]{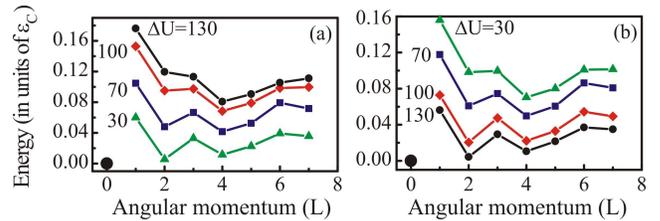}\end{center}
\vspace*{-0.5 cm}
\caption{Low-energy excitation spectra of the $\frac13$-FQHE states 
(eight electrons) in the Landau levels (a) $1^{(+)}_{(2)}$, and (b)
$1^{(+)}_{(1)}$, shown for different values of the bias potential (the 
numbers next to the lines are the values of $\Delta U$ in meV).  
The system is fully spin-polarized. The inter-layer hopping integral 
is set to 30 meV and the magnetic field is 15 Tesla. The flux quanta is 
$2S = 21$. The solid dot at $L=0$ depicts the ground state. The energy unit is
$\varepsilon^{}_c=e^2/\kappa\ell_0$.
}
\label{figtwo}
\end{figure}

The collapse of the FQHE gap corresponding to the appearence of anticrossing 
of the $n=1$ Landau levels, is illustrated in Fig.~\ref{figtwo}. The FQHE 
gap has a monotonic dependence on the bias voltage. In the anticrossing region 
the gap disappears for the lower $n=1$ Landau level (Fig.~\ref{figtwo}a) and 
reappears for the higher $n=1$ Landau level (Fig.~\ref{figtwo}b). The evolution 
of the energy spectra of the incompressible liquid is found to be similar for 
other filling factors (such as $\nu =\frac25$ \cite{vadim_unpub}). This behavior 
was never before observed in the FQHE of conventional two-dimensional electron systems.

\begin{figure}
\begin{center}
\includegraphics[width=8.4cm]{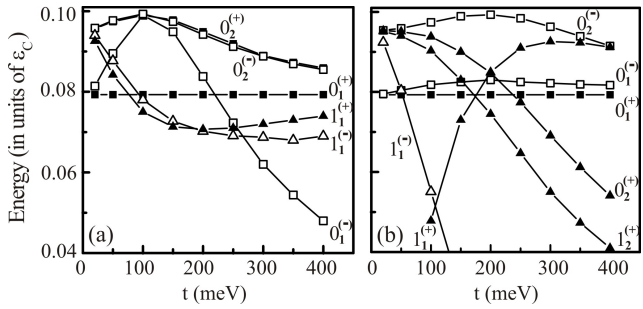}
\end{center}
\vspace*{-0.5 cm}
\caption{The FQHE gaps are shown for different Landau levels.
The labels next to the lines correspond to the labeling of Landau
levels shown in Fig.~\ref{figone}. $\nu=\frac13$-FQHE (eight 
electron) for (a) $\Delta U = 10$ meV, and (b) $\Delta U=300$ meV.
All systems are fully spin polarized. 
}
\label{figthree}
\end{figure}

The strength of the FQHE, i.e., the magnitude of the excitation gap, 
depends on the bias voltage and the inter-layer hopping integral. In 
Fig.~\ref{figthree}, this dependence is shown for $\frac13$-FQHE 
in different Landau levels as a function of $t$. In accordance with 
the properties of Haldane pseudopotentials, the excitation gap of the 
$0_1^{(+)}$ Landau levels does not depend on the bias voltage and on the 
inter-layer hopping integral. The corresponding gap remains constant and 
is equal to the gap of the FQHE in a single layer of graphene in the $n=0$ 
Landau level. For $t=0$ the two layers of graphene are decoupled and 
the bilayer system becomes identical to a monolayer with additional double 
degeneracy. This property is clearly seen in Fig.~\ref{figthree}, where 
for $t=0$ there are only two doubly degenerate FQHE gaps,
corresponding to $n=0$ and $n=1$ single layer Landau levels.

At the zero bias voltage the system has two-fold valley degeneracy, 
which is lifted at finite values of $\Delta U$. At a small bias voltage, 
$\Delta U = 10$ meV, the levels belonging to two valleys are almost 
degenerate, which results in almost the same FQHE gaps of the corresponding 
levels. At the same time the FQHE gaps of $0_1^{(+)}$ and $0_1^{(-)}$ 
levels, which are degenerate at the zero bias voltage, are different.
The origin of these levels is the following: At the zero bias voltage 
there is a four-fold degenerate Landau level with zero energy ($0_1^{(+)}$, 
$0_1^{(-)}$, $0_{-1}^{(+)}$, and $0_{-1}^{(-)}$). At a finite bias voltage,
two of the levels have positive energies (shown in Fig.~\ref{figone}) and the 
other two levels have negative energies. At small values of $\Delta U$, the 
wavefunctions corresponding to the levels $0_1^{(+)}$ and $0_{-1}^{(+)}$ 
of valley $K$ have the form $(0,0,0,\phi^{}_0)$ and $(\phi^{}_0, 0, 0, (t/2^{\frac12})
\phi^{}_1)$, respectively. Here $\phi^{}_n$ are $n$-th `nonrelativistic' Landau 
functions and $t$ is in units of $\hbar v^{}_F/\ell^{}_0$. The corresponding form 
factors $F(q)$, are $L_0$ for level $0_1^{(+)}$ and $\frac{L_0 + (t^2/2) 
L_1}{1+t^2/2}$ for level $0_1^{(-)}$. Although the energies of these levels 
are almost the same the form factors and hence the gaps are 
quite different. In Fig.~\ref{figthree}(a) this difference is clearly visible.
The dependence of the gap of FQHE of level $0_1^{(-)}$ on parameter $t$ 
is nonmonotonic. At $t=0$ the form factor of level $0_1^{(-)}$ is $L_0$ 
and the FQHE gap is exactly equal to the FQHE gap in the $n=0$ Landau 
level of a single graphene layer. At $t=2^{\frac12}\hbar v^{}_F/\ell^{}_0$ the form 
factor is $\frac{L_0+L_1}2$ and the FQHE gap is equal to that in the
$n=1$ Landau level of a single graphene layer. This point corresponds 
to the maximum in Fig.~\ref{figthree}(a). At a large bias voltage, $\Delta U = 300$ 
meV (Fig.~\ref{figthree}b), the FQHE gaps show mainly monotonic dependence 
on the hopping integral. The FQHE gaps of the levels $0_1^{(+)}$ and 
$0_1^{(-)}$, which were quite different in Fig.~\ref{figthree}(a), are now close. 
There is also disappearance of the FQHE in some $n=1$ Landau levels, 
which corresponds to the anti-crossing behavior in Fig.~\ref{figone}. Similar
results are also found for the filling factor $\nu=\frac25$ (i.e., FQHE $\leftrightarrow$ 
non-FQHE transitions occur at the same Landau levels and at 
similar values of $\Delta U$) \cite{vadim_unpub}.

To summarize, we have clearly demonstrated that bilayer graphene
in a strong perpendicular magnetic field reveals some unique properties,
which could allow novel transitions from the FQHE state to a vanishing FQHE
state. These transitions occur within the same Landau level by varying
the bias voltage, i.e, the potential difference between the layers.
Similarly, we have shown that our work on bilayer graphene also results in
new physics: The transitions FQHE $\leftrightarrow$ zero-FQHE, which for
weak inter-layer coupling can be explained as the result of anti-crossing of two
levels, also persists in the limit of strong coupling, where all levels
are strongly coupled. We have established here that there is a fundamental difference
between the two regimes of weak and strong coupling in bilayer graphene.
The boundary between the two regions is determined by the dimensionless parameter
$[t/(\hbar v^{}_F/\ell_0)] \sim 1.5.$ 

We wish to thank David Abergel and Julia Berashevich for very helpful 
discussions. The work has been supported by the Canada Research Chairs 
Program.


\begin{thebibliography}{99}
\bibitem[\ddag]{byline} Electronic address:
tapash@physics.umanitoba.ca

\bibitem{novoselov_kim}
K.S. Novoselov, {\it et al.}, Nature {\bf 438}, 197 (2005);
Y. Zhang, {\it et al.},
{\it ibid.} {\bf 438}, 201 (2005).

\bibitem{review}
D.S.L. Abergel, V. Apalkov, J. Berashevich, K. Ziegler, and 
T. Chakraborty, Adv. Phys. (in press) (2010).

\bibitem{wallace}
P.R. Wallace, Phys. Rev. {\bf 71}, 622 (1947).

\bibitem{mcclure}
J.W. McClure, Phys. Rev. {\bf 104}, 666 (1956).

\bibitem{novoselov_bi}
K.S. Novoselov, {\it et al.}, Nat. Phys. {\bf 2}, 177 (2006);
E. McCann and V. Falko, Phys. Rev. Lett. {\bf 96}, 086805 (2006);
E. McCann, Phys. Rev. B {\bf 74}, 161403 (2006);
T. Ohta, {\it et al.},
Science {\bf 313}, 951 (2006); E.V. Castro, {\it et al.}, Phys. Rev. Lett.
{\bf 99}, 216802 (2007); M. Koshino and E. McCann, Phys. Rev. B {\bf 81},
115315 (2010).

\bibitem{pereira}
J.M. Pereira, Jr., F.M. Peeters, and P. Vasilopoulos,
Phys. Rev. B {\bf 76}, 115419 (2007).

\bibitem{abergel}
D.S.L. Abergel and T. Chakraborty, Phys. Rev. Lett. {\bf 102}, 056807 (2009).

\bibitem{fqhe_book} T. Chakraborty, and P. Pietil\"ainen, {\it The
Quantum Hall Effects} (Springer, New York, 1995), 2nd edition;
T. Chakraborty, Adv. Phys. {\bf 49}, 959 (2000).

\bibitem{stormer}
D.C. Tsui, H.L. St\"ormer, and A.C. Gossard, Phys. Rev. Lett. {\bf 48}, 1559 (1982);
R.B. Laughlin, {\it ibid.} {\bf 50}, 1395 (1983).

\bibitem{vadim_fqhe}
V.M. Apalkov and T. Chakraborty, Phys. Rev. Lett. {\bf 97}, 126801 (2006).

\bibitem{fqhe_kim}
K.I. Bolotin, {\it et al.},
Nature {\bf 462}, 196 (2009); see also, X. Du, et al., {\it ibid.}
{\bf 462}, 192 (2009).

\bibitem{nakamura}
M. Nakamura, E.V. Castro, and B. Dora, Phys. Rev. Lett. {\bf 103}, 266804 (2009);
Y. Zhao, {\it et al.},
{\it ibid.} {\bf 104},
66801 (2010).

\bibitem{haldane}
F.D.M. Haldane, Phys. Rev. Lett. {\bf 51}, 605 (1983).

\bibitem{goerbig}
M.O. Goerbig, R. Moessner, and B. Doucot,Phys. Rev. B {\bf 74} 161407(R) (2006).

\bibitem{fano}
G. Fano, F. Ortolani, and E. Colombo, Phys. Rev. B {\bf 34}, 2670 (1986).


\bibitem{vadim_unpub}
V. Apalkov and T. Chakraborty, unpublished (2010).

\end{thebibliography}
\end{document}